\documentclass[aps,twocolumn,prl,amsmath,amssymb,notitlepage,superscriptaddress,longbibliography,nofootinbib,longbibliography]{revtex4-2}

\usepackage{graphicx}% Include figure files
\usepackage{dcolumn} % Align table columns on decimal point
\usepackage{bm} % bold math
\usepackage{hyperref}% add hypertext capabilities
\usepackage[mathlines]{lineno}% Enable numbering of text and display math
\usepackage{xcolor}
\definecolor{darkblue}{rgb}{0,0,0.6}
\definecolor{darkred}{rgb}{0.6,0,0}
\definecolor{darkgreen}{rgb}{0,0.6,0}
\definecolor{bleu}{rgb}{0,0.44,0.72}
\usepackage{setspace}
\usepackage{ulem}
\usepackage{enumerate}
\usepackage{tikz}
\newcommand{\cF}{\mathcal{F}}

\newcommand{\cL}{\mathcal{L}}

\newcommand\rmd{\mathrm{d}}

\newcommand\bfr{{\bf r}}

\newcommand\bfv{{\bf v}}

\newcommand{\sigmaea}{\sigma^{\rm e,a}}

\begin{document}

\title{Active topological strings in renewing nematopolar fluids}

\author{Alberto Dinelli}
\email{alberto.dinelli@unige.ch}
\affiliation{Department of Biochemistry, University of Geneva, 1211 Geneva, Switzerland}
\affiliation{Department of Theoretical Physics, University of Geneva, 1211 Geneva, Switzerland}
\author{Ludovic Dumoulin}
\email{ludovic.dumoulin@unige.ch}
\affiliation{Department of Biochemistry, University of Geneva, 1211 Geneva, Switzerland}
\affiliation{Department of Theoretical Physics, University of Geneva, 1211 Geneva, Switzerland}
\author{Karsten Kruse}
\email{karsten.kruse@unige.ch}
\affiliation{Department of Biochemistry, University of Geneva, 1211 Geneva, Switzerland}
\affiliation{Department of Theoretical Physics, University of Geneva, 1211 Geneva, Switzerland}

\date{\today}

\begin{abstract} 
Active matter often simultaneously exhibits different kinds of orientational order and, in many cases of biological interest, 
undergoes continuous material renewal.
%its constituents are continuously renewed. 
In renewing nematopolar fluids we find stable topological strings, structures consisting of two nematic point defects connected by a defect line in the polar field. We identify the mechanism underlying string stabilization and unveil how string length is determined. In the presence of active stress, we observe active-string chaos. Our work identifies continuous material renewal as a generic mechanism underlying the stabilization of topological defect structures in systems with mixed order parameters. It could be used for orchestrating living matter during development and other biological processes.
\end{abstract}

\maketitle

%\section{Introduction}
Topological defects are configurations of a continuous order-parameter field that cannot be  continuously deformed into a fully ordered state. Such defects appear in numerous physical systems and attract sustained interest in the field of active matter~\cite{shankar2022topo}. In particular, they play a central role in the emergence of nonequilibrium phases~\cite{ramaswamy2003active,Sanchez:2012gt,thampi2014active,doostmohammadi2018active}, can be used to measure phenomenological parameters of active matter~\cite{Zhang:2018ha,Blanch-Mercader.2021lzf}, and guide morphological events during animal development~\cite{Maroudas-Sacks.2021,Ravichandran.2025}.

The nature and dynamics of topological defects are constrained by the symmetries of the order parameter. For example, point defects in two-dimensional systems are characterized by their topological charge, that is, the number of turns the order parameter makes along a closed curved around the defect. For a polar order parameter, breaking head-tail symmetry, only integer topological charges are possible. In contrast, point defects in a nematic order parameter exhibiting head-tail symmetry can have half-integer topological charges. In the presence of active stress, $+1/2$ nematic defects are themselves polar and self-propel~\cite{Sanchez:2012gt,giomi2014defect,Saw:2017gn,doostmohammadi2018active}.

Interestingly, the description of some systems requires multiple order parameters. For example, ferroelectric nematic liquid crystals~\cite{chen2020first,lavrentovich2020ferroelectric,sebastian2022ferroelectric} simultaneously exhibit nematic and polar order. Importantly, this is also true for cytoskeletal structures that are composed of filamentous polymers and vital for living cells. Among those structures are stress fibers through which cells exert stress on their environment, lamellipodia that are involved in cell migration~\cite{svitkina1997analysis}, and potentially the cleavage furrow during cell division~\cite{Fishkind1993division}. Tissues have been suggested to simultaneously exhibit nematic and hexatic order~\cite{Armengol-Collado.2023}, but see also Ref.~\cite{happel2025}.

New structures can appear when different order parameters interact. For example,  alignment between nematic and polar order induces elongated polar defects with integer charge, called topological strings~\cite{lee1985strings,carpenter1989phase,vafa2025phase,mishra2025string}. In equilibrium systems, such topological strings are transient and eventually annihilate in pairs unless enforced by boundary conditions~\cite{mishra2025string}.

Here, we study the emergence and stabilization of topological defects in a compressible nematopolar fluid with material renewal, that is, with non-conserved mass. 
%It is motivated by the cytoskeleton, where renewal corresponds to continuous filament assembly and disassembly. 
In biological systems, renewal arises, for example, via birth-and-death dynamics in populations of cells~\cite{toner2012birth,bunin2017ecological,hallatschek2023proliferating} or via continuous assembly and disassembly of filaments in the cytoskeleton~\cite{blanchoin2014actin}.
We show how renewal maintains hydrodynamic forces that can stabilize energetically-unfavorable configurations for isolated strings. This mechanism leads to a rich variety of stationary configurations with multiple coexisting defects. Furthermore, by elucidating the charge-dependent nature of these forces, we show how the steady-state length of isolated strings differs between strings of different charges. Although active stress alone does not stabilize topological strings, it reorients neighboring stable strings in the presence of renewal and selects specific types of strings with topological charge +1. Our findings suggest the existence of stable complex defect structures in active matter with renewal that could be the basis for mechanical organization of biological processes.

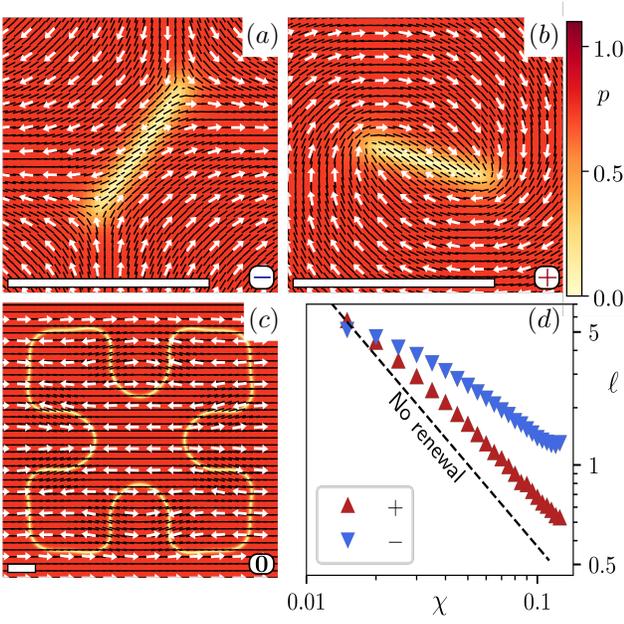
\begin{figure}
    \centering
    \begin{tikzpicture}
    \def\w{15pt}
    \def\h{15pt}
        \path (0,0) node [anchor=north west, text width=1\columnwidth] {\includegraphics[width=1\columnwidth]{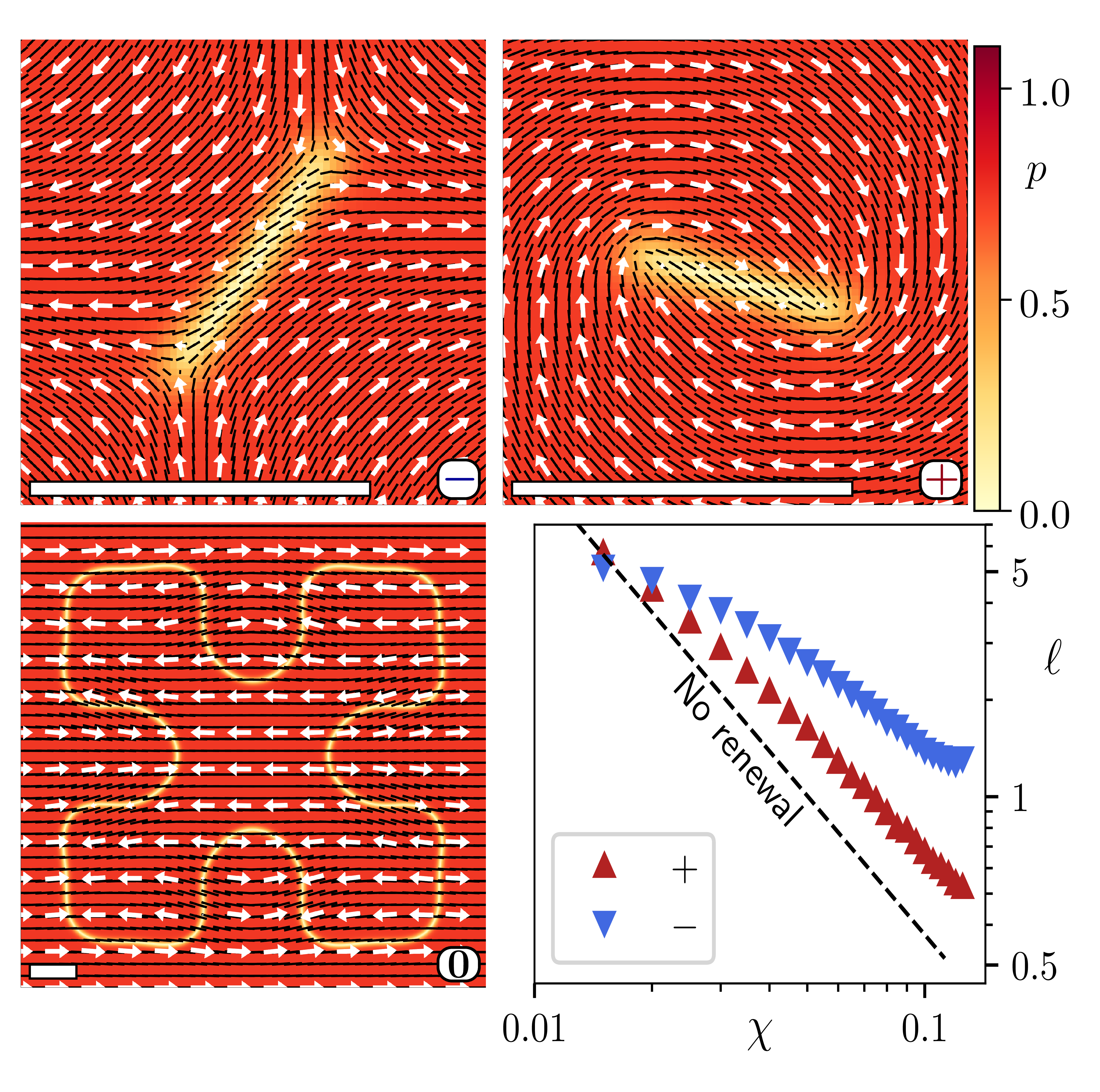}};
        \path (106.5pt,-19.5pt) node [inner sep=0pt, draw=white, rectangle, align=center, fill=white, minimum height=\h, minimum width=\w] {$(a)$};
        \path (213pt,-19.5pt) node [inner sep=0pt, draw=white, rectangle, align=center, fill=white, minimum height=\h, minimum width=\w] {$(b)$};
        \path (107pt,-127pt) node [inner sep=0pt, draw=white, rectangle, align=center, fill=white, minimum height=\h, minimum width=\w] {$(c)$};
        \path (213pt,-127pt) node [inner sep=0pt, align=center, rectangle,  minimum height=\h, minimum width=\w] {$(d)$};
    \end{tikzpicture}    
    \caption{Renewal stabilizes topological strings. \textbf{(a-c)} Heatmaps of the polar order parameter $p$ for topological strings of charge $-1$ (a), $+1$ (b), and $0$ (c) in the presence of material renewal. White arrows: polar director, black dashes: nematic director, scale bar: unit length. \textbf{(d)}
    Steady-state length $\ell$ of $\pm1$ strings ($+$/$-$) as a function of the nematopolar coupling strength $\chi$.
   Parameters: $\rho_0=0.6$, $\tau^{-1}=5$, and $\chi=0.1$ (a-c).}
    \label{fig:single-defect}
\end{figure}

\textit{Hydrodynamics of nematopolar fluids.---}We study a compressible nematopolar fluid in two dimensions. The material density $\rho$ evolves according to~\cite{Hannezo:2015ba,dumoulin2025defect} 
\begin{align}
\partial_t \rho +\partial_\alpha (\rho v_\alpha - D \partial_\alpha \rho)&=  - \tau^{-1} (\rho-\rho_0),\label{eq:dyn-rho}
\end{align}
where $v_\alpha$ with $\alpha=x,y$ are the components of the fluid's velocity field and $D$ a diffusion constant. The diffusion term captures stochastic effects in the fluid dynamics and prevents instabilities on small length scales. We implicitly sum over repeated indices. The right-hand term describes material renewal, relaxing the density to its steady-state value $\rho_0$ at rate $\tau^{-1}$. More involved expressions are possible~\cite{Hannezo:2015ba,barberi2023localized} and can lead to interesting dynamics not captured by Eq.~\eqref{eq:dyn-rho}. In the present, work we focus on the generic impact of renewal and leave the analysis of nonlinear renewal for future studies.

The nematopolar order is captured by the polar field $\mathbf{p}$ and the traceless nematic tensor $\mathsf{Q}$. We write $p_\alpha=p\,\hat{p}_\alpha$ and $Q_{\alpha\beta}=S\left(\hat{n}_\alpha \hat{n}_\beta-\tfrac{1}{2}\delta_{\alpha\beta}\right)$, where $\hat{\mathbf{p}}$ and $\hat{\mathbf{n}}$ are the polar and nematic directors with $\hat{p}_\alpha^2=\hat{n}_\alpha^2=1$. The polar and nematic order parameters are given by $p^2=p_\alpha p_\alpha$ and $S^2=2 Q_{\alpha\beta}Q_{\beta\alpha}\equiv2Q^2$. For the associated free energy $\cF = \int f \rmd \bfr$, we choose
\begin{align}
  f = &\rho^2\left[ -\frac{\alpha}{2}\left(\frac{\rho}{\rho_0}- Q^2\right)Q^2 + \frac{\kappa_Q}{2} (\partial_\alpha Q_{\beta\gamma} )(\partial_\alpha Q_{\beta\gamma})  
  \right. \nonumber\\
  & \left. + \;\>\frac{\beta}{4} p^4 + \frac{\kappa_p}{2} (\partial_\alpha p_\beta )(\partial_\alpha p_\beta ) - \frac{\chi}{2} Q_{\alpha\beta} p_\alpha p_\beta\right]  + \frac{a}{4} \rho^4 . 
  \label{eq:free_energy}
\end{align}

The polar and nematic order parameters are governed by Landau-de Gennes terms, where we have dropped the quadratic term for $p$ and chosen $\alpha,\beta>0$. Thus, for the polar field, the disordered state with $p=0$ is energetically favored, whereas the system tends to be nematically ordered with $S^2=\rho/\rho_0$. The nematopolar coupling modifies these values, such that the free energy is minimized for $S^\star=\sqrt{1+\chi^2/(4 \alpha\beta)}$ and $p^\star=\sqrt{|\chi| S^\star/(2 \beta)}$ if $\rho=\rho_0$. 

For the corresponding Frank free energy, we take the one-constant approximation, such that splay and bend deformations of the polar and the nematic fields are associated with the same respective elastic constants $\kappa_p$ and $\kappa_Q$. The coupling term between the two kinds of orientational order tends to align the two directors if $\chi>0$. They tend to adopt a locally perpendicular configuration in the opposite case. Finally, terms containing $\mathsf{Q}$ or $\mathbf{p}$ are scaled with $\rho^2$, so that their energetic contributions increase with higher fluid density. Steric interactions between the fluid's constituents eventually limit its density, which is captured by the fourth order term in $\rho$ with $a>0$.

The orientational order fields evolve as
\begin{align}
    \label{eq:dyn-p}
    \partial_t p_\alpha  + v_\beta \partial_\beta p_\alpha + \omega_{\alpha\beta} p_\beta &= \gamma^{-1} h_\alpha \\
     \partial_t Q_{\alpha\beta}   +v_\gamma \partial_\gamma Q_{\alpha\beta}  + \omega_{\alpha\gamma} Q_{\gamma\beta} &=  \Gamma^{-1} H_{\alpha\beta} .
\label{eq:dyn-Q}
\end{align}
On the left-hand sides, we have the convective co-rotational derivative of $\mathbf{p}$ and $\mathsf{Q}$, where $\omega_{\alpha\beta}  = \frac{1}{2} [\partial_\alpha v_\beta - \partial_\beta v_\alpha]$ denote the components of the vorticity tensor. The fields $\mathbf{h}\equiv-\delta\mathcal{F}/\delta\mathbf{p}$ and $\mathsf{H}\equiv-\delta\mathcal{F}/\delta\mathsf{Q}$ denote, respectively, the polar and nematic molecular fields and $\gamma$ and $\Gamma$ are the associated rotational viscosities. For simplicity, we neglect the effect of flow alignment.

The velocity field is determined by momentum conservation. Since we focus on the limit of vanishing Reynolds number, it takes the form of force balance. We consider the fluid to be coupled to a substrate with friction coefficient $\xi$. Consequently, we have 
\begin{align}
\partial_\beta \sigma^\mathrm{tot}_{\alpha\beta} = \xi v_\alpha, 
\end{align}
where $ \mathsf{\sigma}^\mathrm{tot}$ is the total stress tensor~\cite{leslie1968some},
\begin{align}
\label{eq:stress}
\mathsf{\sigma}^\mathrm{tot} = & \mathsf{\sigma}^\mathrm{visc} + \mathsf{\sigma}^\mathrm{e}.
\end{align}
The viscous stress has components $\sigma^\mathrm{visc}_{\alpha\beta}=\eta(\partial_\alpha v_\beta+\partial_\beta v_{\alpha})$ with $\eta$ being the viscosity. The Ericksen stress $\mathsf{\sigma}^\mathrm{e}$ generalizes the hydrostatic pressure $\Pi\equiv\rho \partial_\rho f-f$ to account for molecular anisotropies~\cite{ericksen1962hydrostatic,leslie1968some,de1995physics}. Its components are given in the Supplemental Material (SM)~\cite{supp}. 

We adopt periodic boundary conditions and use nondimensionalized parameters as detailed in the End Matter. Furthermore, for simplicity, we set the nondimensionalized parameters $\gamma=\Gamma=1$. Numerical solutions were performed on a square domain of size $L^2$ and on a grid with $N^2$ points. The numerical methods and the initial conditions are specified in the SM~\cite{supp}. 

\textit{Renewal stabilizes topological strings.---}We now turn our attention to topological defects in the nematic-order field, which can have half-integer charges. Since defects of the polar order field can only have integer charges, the system gets frustrated in presence of half-integer nematic point defects if polar and nematic order are coupled, $\chi\neq0$. This frustration is resolved by creating topological strings, that is, lines along which $p=0$ connecting nematic point defects~\cite{vafa2025phase,mishra2025string}. Across these lines, the polar field changes sign. The topological charge associated with such strings is given by the sum of the charges of the two extreme nematic point defects~\cite{supp}.

The energetics of topological strings is determined by the Coulomb interaction between the nematic point defects and the energetic cost for $p=0$ along the string, which induces an effective line tension~\cite{supp,vafa2025phase}. For zero-charge strings, both forces are attractive, such that these strings self-annihilate in the absence of renewal. In contrast, for topological strings with charge $\pm1$ ($\pm1$ strings) the Coulomb repulsion of the nematic point defects opposes the attractive line tension. Sufficiently isolated strings adopt a length $\ell$ that is determined by balancing these two forces, until they eventually annihilate by fusing with oppositely charged strings, see SM Movie~1~\cite{supp}. As the line tension increases as $\chi^{3/2}$, the length decreases with the nematopolar coupling strength as $\chi^{-3/2}$~\cite{supp}.

This scenario changes dramatically in the presence of renewal, $\tau^{-1} > 0$. First, $\pm1$ strings exist at steady-state, Fig.~\ref{fig:single-defect}a,b, SM Movie~2~\cite{supp}. The orientation of the nematic point defects at the extremes of a string is not fixed by system parameters but established spontaneously. For $+1$ strings, the extreme nematic point defects minimize elastic distortions by orienting in an antiparallel way. The far polar field of such strings is thus typically a spiral. In the form of loops, also topological strings with zero charge can be stable, Fig.~\ref{fig:single-defect}c, SM Movie~3~\cite{supp}. 

The steady-state string length $\ell$ deviates from the value for $\tau^{-1} = 0$, with $-1$ strings being longer than those with charge $+1$, Fig.~\ref{fig:single-defect}d. These results show that, in the presence of renewal, there are additional charge-dependent forces determining the behavior of topological strings. Note that also the functional dependence of $\ell$ on the parameter $\chi$ is different than for $\tau^{-1} = 0$. 

\begin{figure}
    \centering
    \begin{tikzpicture}
    \def\w{15pt}
    \def\h{15pt}
        \path (0,0) node [anchor=north west, align=left, text width=\columnwidth] {\includegraphics[width=\columnwidth]{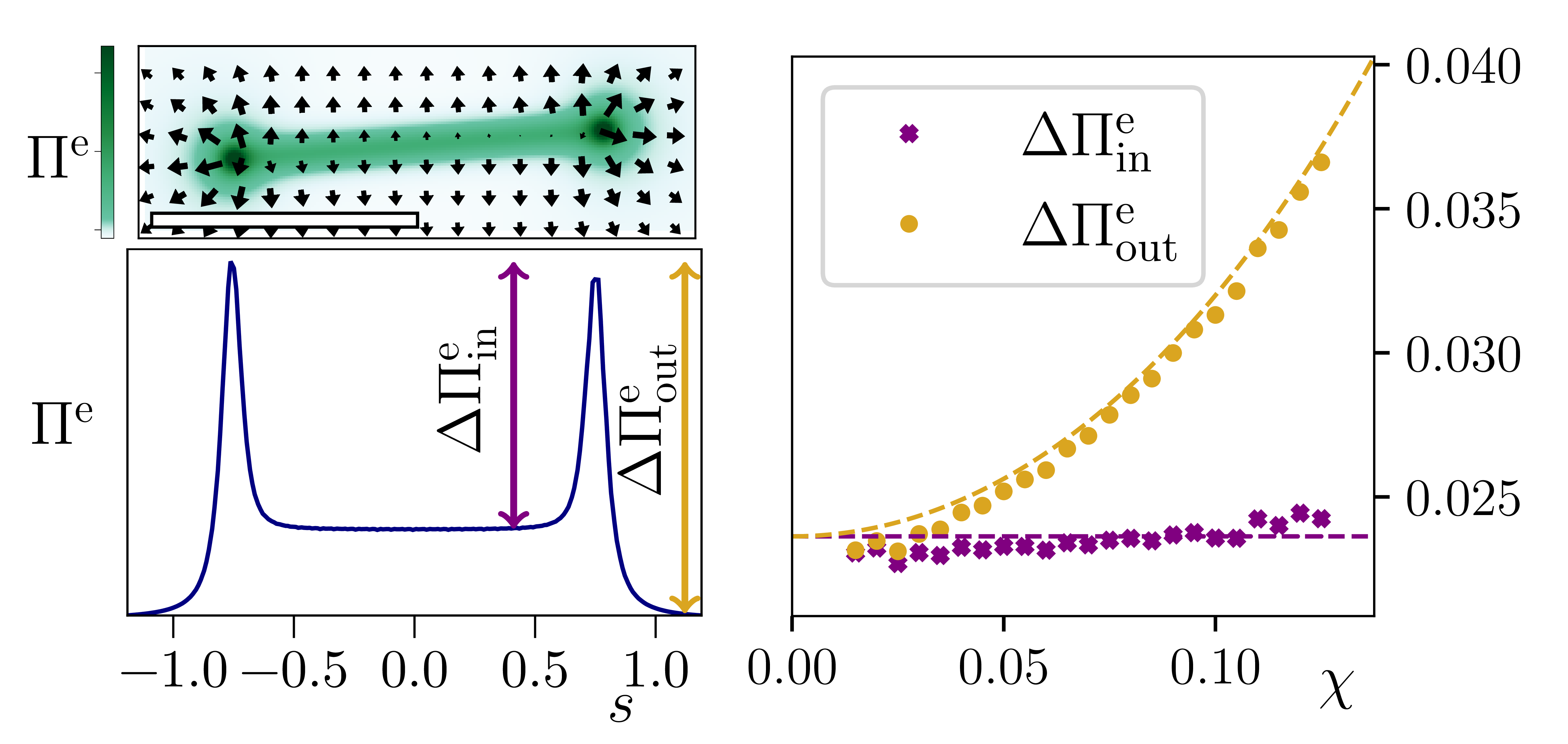}};
        \path (5pt,-15pt) node [inner sep=0pt, draw=white, rectangle, align=center, minimum height=\h, minimum width=\w] {$(a)$};
        \path (215pt,-15.5pt) node [inner sep=0pt, draw=white, rectangle, align=center, fill=white, minimum height=\h, minimum width=\w,opacity=1] {$(b)$};
    \end{tikzpicture}    
    \caption{Differences in effective pressure $\Pi^{\rm e}$ drive flows stabilizing topological strings. \textbf{(a)} Effective pressure $\Pi^{\rm e}$ around a topological string. Top panel: Heatmap of $\Pi^{\rm e}$ for a $-1$ string in the presence of renewal. Black arrows: velocity field $\mathbf{v}$, scale bar: unit length. Bottom panel: Profile of $\Pi^{\rm e}$ as a function of the coordinate $s$ along the string. Pressure differences $\Delta \Pi^{\rm e}_\mathrm{in}$ and $\Delta\Pi^{\rm e}_\mathrm{out}$ are, respectively, taken between the nematic point defect and the string interior or the defect-free bulk. \textbf{(b)} Pressure differences as a function of the nematopolar coupling parameter $\chi$. Dashed lines: analytical estimates, see text. Parameters as in Fig.~\ref{fig:single-defect}. }
    \label{fig:pressure}
\end{figure}

In the presence of renewal, the fusion of topological strings with opposite charges as well as the collapse of strings with vanishing charge is prevented by flows emanating from the strings, Fig.~\ref{fig:pressure}a. To understand the origin of these flows, we introduce the effective pressure $\Pi^{\rm e} := \frac{1}{2} \text{Tr} [\sigma^{\rm e}]$, which extends the hydrostatic pressure $\Pi$ and includes the isotropic contributions due to elastic stresses. To first approximation, flows can be captured by gradients in the effective pressure through Darcy-Brinkman's law, $v_\alpha \simeq - G_{\alpha\beta} \ast\partial_\beta \Pi^{\rm e}$. Here, $G_{\alpha\beta}(\bfr)$ is the Green function associated to the frictional-viscous flow and $\ast$ denotes the convolution product~\cite{brinkman1949calculation,supp}, see End Matter.
The Green function $G_{\alpha\beta}$ propagates local forces up to distances of the order of the hydrodynamic length $\ell_h=\sqrt{\eta/\xi} = 1$~\cite{supp}. At the same time, as fluid is continuously advected out of the string region, renewal tends to refill it with new material and thereby sustains these flows~\cite{dumoulin2025defect}.

The effective pressure is also responsible for increasing the length of topological strings compared to the case without renewal. Indeed, between the two extreme nematic point defects of a string, $\Pi^\mathrm{e}$ is lower than in the point defects, Fig.~\ref{fig:pressure}a. However, along the string, $\Pi^\mathrm{e}$ is also larger than in the region surrounding the string. As a consequence, there is a net flow tending to separate the nematic point defects by advection. Note that this flow is  driven by local pressure gradients associated with the extreme nematic point defects. As a consequence, it is essentially independent of the string length.

To analyze how the flows depend on the nematopolar coupling constant $\chi$, we estimate the difference in effective pressure between the nematic point defect and the string center or the surrounding region. For the chosen parameter values, density gradients are small, and we neglect their contribution in our analysis. Furthermore, we neglect contributions from the Frank elastic terms, so that $\Pi^{\rm e}\simeq\Pi$ in the absence of deformations. %Then, the free energy outside a string is essentially zero. 
Along the line connecting the two extreme nematic point defects, we have $S=1$ and $p=0$, whereas $S=p=0$ inside them. Consequently,
\begin{align}
    \Delta\Pi^{\rm e}_\mathrm{out} &= \Delta\Pi^{\rm e}_\mathrm{in}  \left(1 + \frac{\chi^2}{4 \alpha \beta}\right) \left(1 + \frac{\chi^2}{12 \alpha\beta}\right),
    \label{eq:grad_press_2}
\end{align}
where $\Delta\Pi^{\rm e}_\mathrm{out}$ and $\Delta\Pi^{\rm e}_\mathrm{in}$ are the differences in hydrostatic pressure between a point defect and the surrounding region or the interior of the string, Fig.~\ref{fig:pressure}a. The same analysis shows that $\Delta\Pi^{\rm e}_\mathrm{in}$ is constant. Using the above equation and %the expression for 
fitting the value of $\Delta\Pi^{\rm e}_\mathrm{in}$, we see that our estimate agrees well with our numerical results, Fig.~\ref{fig:pressure}b.

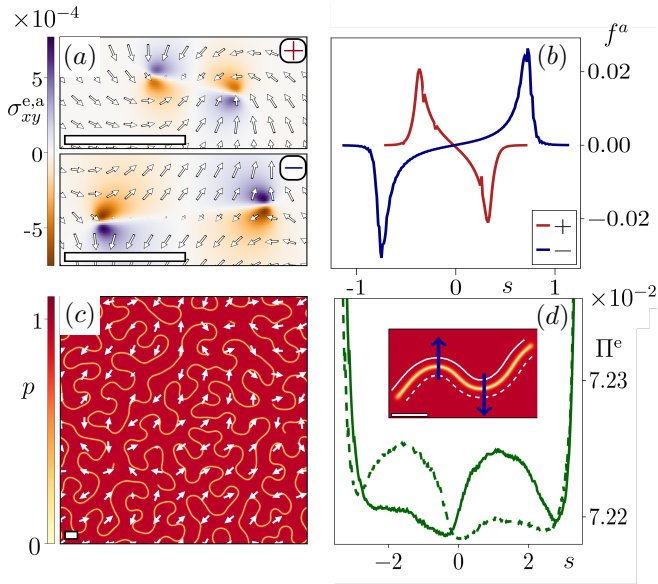
\begin{figure}
    \begin{tikzpicture}
    \def\w{15pt}
    \def\h{15pt}
        \path (0,0) node [inner sep=0, anchor=north west, align=left, text width=1\columnwidth] {\includegraphics[width=\columnwidth,trim={70 0 50 0},clip]{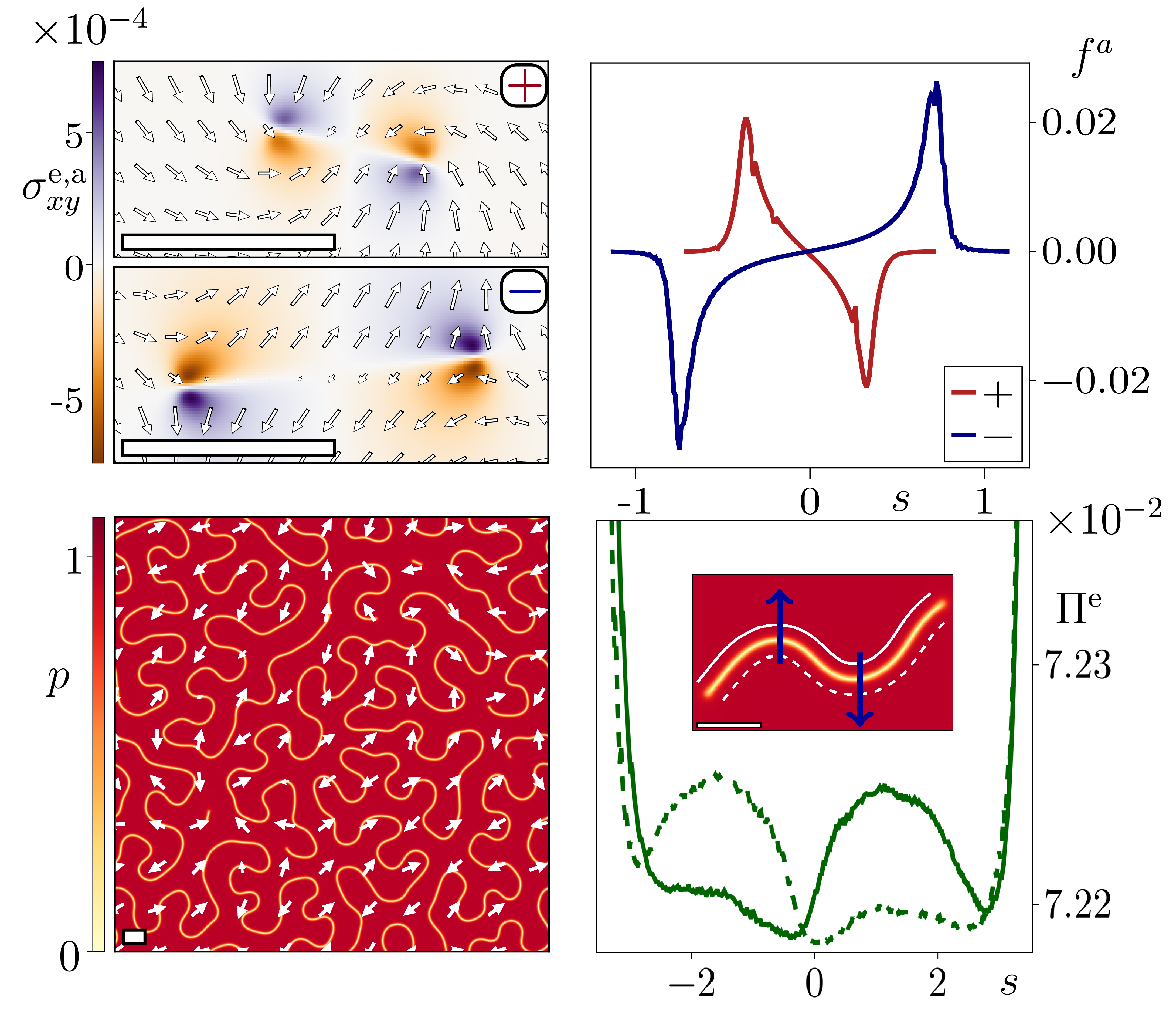}};
        \path (28.3pt,-20.9pt) node [inner sep=0pt, fill=white, rectangle, align=center, minimum height=\h, minimum width=\w] {$(a)$};
        \path (28.3pt,-118.59pt) node [inner sep=0pt, fill=white, rectangle, align=center, minimum height=\h, minimum width=\w] {$(c)$};
        \path (207pt,-21pt) node [inner sep=0pt, align=center, minimum height=\h, minimum width=\w,opacity=1] {$(b)$};
        \path (207pt,-119.5pt) node [inner sep=0pt, align=center, minimum height=\h, minimum width=\w,opacity=1] {$(d)$};
    \end{tikzpicture}    
    \caption{Charge-dependent hydrodynamic forces control the length and stability of topological strings. \textbf{(a)} Heatmap of the antisymmetric part of the Ericksen stress $\sigmaea_{xy}$ for $\pm1$ strings ($+$/$-$). Arrows: polar field $\mathbf{p}$. \textbf{(b)} Longitudinal forces resulting from $\mathsf{\sigma}^\mathrm{e,a}$ as a function of the coordinate $s$ along $\pm1$ strings (red/blue). \textbf{(c)} Heatmap of the polar order parameter $p$ for an unstable $-1$ topological string. White arrows: $\mathbf{p}$. \textbf{(d)} Effective pressure $\Pi^{\rm e}$ adjacent to an unstable $-1$ topological string. Inset: zoom on unstable wiggling string. White lines indicate where profiles of $\Pi^{\rm e}$ were measured (line styles match). Arrows indicate net forces resulting from pressure imbalance. Parameter values: As in Fig.~\ref{fig:single-defect} and $\chi=0.1$ (a,b) and $\chi=0.15$ (c,d). Scale bars: unit length.} 
    \label{fig:charge_dependent}
\end{figure}

\textit{Renewal maintains charge-dependent forces on topological strings.---}The effective pressure difference Eq.~\ref{eq:grad_press_2} does not depend on the charge of the nematic point defects and can thus not account for the differences in length of strings with opposite charges, Fig.~\ref{fig:single-defect}d. Beyond the effective pressure, this holds for the symmetric part of the Ericksen stress. This is readily seen when expressing its components in terms of the angles $\nu$ and $\psi$ formed by the directors $\hat{\mathbf{p}}$ and $\hat{\mathbf{n}}$ with the $x$-axis. Under the transformations $\nu\to-\nu$ and $\psi\to-\psi$ the charge of a point defect is inverted, however, the symmetric part of the Ericksen stress remains invariant. 

We thus turn to the antisymmetric part $\mathsf{\sigma}^\mathrm{e,a}$ of the Ericksen stress, where $\sigma^\mathrm{e,a}_{\alpha\beta}=\tfrac{1}{2}\left(p_\alpha h_\beta-p_\beta h_\alpha\right) + H_{\alpha\gamma}Q_{\gamma\beta}-H_{\beta\gamma}Q_{\gamma\alpha}$. We decompose its only nonvanishing component $\sigma^\mathrm{e,a}_{xy}$ into an elastic part $\sigma^\mathrm{el}$ depending on gradients of the polar and nematic fields, and a misalignment part $\sigma^\mathrm{mis}$, which vanishes if $\hat{\mathbf{p}}$ and $\hat{\mathbf{n}}$ are aligned. In the limit where we can neglect density gradients, we have
\begin{align}
    \sigma^\mathrm{el} \simeq & \frac{\kappa_p}{2} \rho_0^2 \left[ \nabla p^2 \cdot \nabla \nu + p^2 \nabla^2 \nu\right] \nonumber \\
    &+
    \kappa_Q \rho_0^2  \left[ \nabla S^2 \cdot \nabla \psi + S^2 \nabla^2 \psi\right].
\end{align}
This term is zero for circularly symmetric point defects, but not for topological strings that break this symmetry. For the misalignment term, we find
\begin{align}
\sigma^\mathrm{mis} = & \frac{\chi}{4}  \rho^2 S p^2\sin  [2 (\nu - \psi)].
\label{eq:sigma_mis}
\end{align}
Both components change sign when $\nu\to-\nu$ and $\psi\to-\psi$ and thus depend on the defect charge. 

In Figure~\ref{fig:charge_dependent}a, we illustrate the charge-dependence of the antisymmetric part of the Ericksen stress. It has a quadrupolar structure and exhibits poles at the string extremes. In the presence of renewal, the misalignment term dominates over the elastic term, SM~\cite{supp}. The force resulting from the antisymmetric part of the Ericksen stress, $\mathbf{f}^a=\nabla\cdot\mathsf{\sigma}^\mathrm{e,a}$, has a component along the string axis. It leads to contraction of topological strings with positive charge and to extension of those with negative charge, Fig.~\ref{fig:charge_dependent}b. As these are the only charge-dependent forces, they alone are responsible for the observed length differences, Fig.~\ref{fig:single-defect}d. 

Charge-dependent forces additionally lead to a transverse instability for $-1$ strings, Fig.~\ref{fig:charge_dependent}c. This instability is linked to the effective pressure, Fig.~\ref{fig:charge_dependent}d. Indeed, on the two opposite sides of an unstable string, $\Pi^{\rm e}$ typically differs, resulting in net transverse forces. In this case, topological strings expand until further growth is blocked by adjacent strings, see SM Movie 4.

\begin{figure}
    \centering
    \begin{tikzpicture}
    \def\w{15pt}
    \def\h{15pt}
        \path (0,0) node [anchor=north west, text width=1\columnwidth] {\includegraphics[width=1\columnwidth]{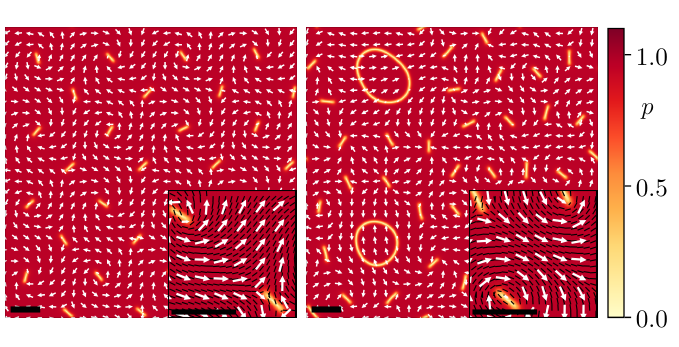}};
        \path (106pt,-21pt) node [inner sep=0pt, draw=white, rectangle, align=center, fill=white, minimum height=\h, minimum width=\w] {$(a)$};
        \path (215pt,-21pt) node [inner sep=0pt, draw=white, rectangle, align=center, fill=white, minimum height=\h, minimum width=\w] {$(b)$};
    \end{tikzpicture}    
    \caption{States of coexisting strings are stabilized by hydrodynamic repulsion. Heatmap of the polar order parameter $p$ for steady-state configurations. ({\bf a}) String lattice for $\tau^{-1}=0.5$. ({\bf b}) Coexistence between integer-charge stings and loops for $\tau^{-1}=2$. Parameter values: $\rho_0=0.8$ and $\chi=0.15$, scale bar: unit length.}
    \label{fig:multiple-defect}
\end{figure}

Hydrodynamic forces maintained by renewal not only affect individual string dynamics, but can also drive the emergence of self-organized lattices of topological strings with alternating charges $+1$ and $-1$, Fig.~\ref{fig:multiple-defect}a and SM Movie 5. In this case, there is no specific pattern in the orientation of polar and nematic fields within each topological string.
In case the initial condition led to the formation of closed strings of charge $0$, the lattice was distorted in their vicinity, Fig.~\ref{fig:multiple-defect}b and SM Movie 6.

\textit{Impact of active stress on topological strings.} Finally, we consider the impact of an active component of the mechanical stress on topological strings. To this end, we extend the expression \eqref{eq:stress} of the stress and write $\mathsf{\sigma}^\mathrm{tot} = \mathsf{\sigma}^\mathrm{visc} + \mathsf{\sigma}^\mathrm{e}+ \mathsf{\sigma}^\mathrm{act}$. For the active stress we take $\sigma^\mathrm{act}_{\alpha\beta}=-\zeta\rho Q_{\alpha\beta}$, such that the fluid is extensile for $\zeta>0$ and contractile otherwise~\cite{dedenon2025noise}.  

\begin{figure}
    \centering
    \centering
    \begin{tikzpicture}
    \def\w{15pt}
    \def\h{15pt}
        \path (0,0) node [anchor=north west, text width=1\columnwidth, inner sep=0] {\includegraphics[width=1\columnwidth,trim={0 0 0 0},clip]{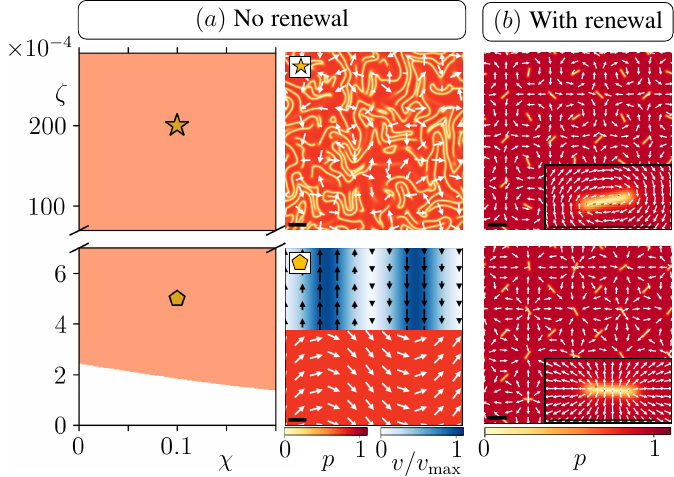}};
    \end{tikzpicture}  
    \caption{Active stress generates spontaneous flows and align nematic point defects with topological strings in nematopolar fluids. \textbf{(a)} Stability boundary of the homogeneous, uniformly orientationally ordered state as a function of the nematopolar coupling parameter $\chi$. Snapshots of numerical solutions for $\zeta=5 \times 10^{-4}$ (pentagon) and $\zeta = 0.02$ (star). Parameters: $\rho_0=0.6$ and $\tau^{-1}=0$. \textbf{(b)} In the presence of renewal, defect lattices with either vortices or lattices are stabilized. Same parameters as in Fig.~\ref{fig:multiple-defect}a except for $\tau^{-1}=5$ and $\zeta=5 \times 10^{-4}$ (top), $\zeta=-5 \times 10^{-4}$ (bottom). Scale bars: unit length.}
    \label{fig:active_phase_diagram}
\end{figure}

We first study the dynamics of an active nematopolar fluid in the absence of renewal. At a critical value of $\zeta>0$ ($\zeta<0$), the steady state with $\rho=\rho_0$, $S_0=S^\star$, $p_0=p^\star$, and $\hat{\mathbf{n}}=\hat{\mathbf{x}}$ becomes unstable through a type-II bend (splay)
instability~\cite{cross1993pattern,nejad2021memory}, with the fastest growing mode having a wave-vector $\mathbf{q}$ directed along $\hat{\mathbf{x}}$ (along $\hat{\mathbf{y}}$), Fig.~\ref{fig:active_phase_diagram}a and SM~\cite{supp}. The nematopolar coupling facilitates this instability. This facilitation results from an increase of the nematic order $S^\star$ with increasing absolute values of the coupling parameter $\chi$. This in turn effectively enhances the active stress.

Close to the onset of the transition, the system exhibits a periodic modulation of the orientation angles $\nu$ and $\psi$ along the $y$-axis. This modulation is accompanied by self-sustained flows, Fig.~\ref{fig:active_phase_diagram}a (bottom). Note that these flows are not due to flow alignment~\cite{Voituriez:2007jy}, which is absent in our system. Upon increasing the active stress further, a chaotic state emerges, which is analogous to chaotic states in active nematics~\cite{ramaswamy2003active,Sanchez:2012gt,thampi2014active}. In this state, dynamic topological strings with charge $0$ proliferate, Fig.~\ref{fig:active_phase_diagram}a (top) and SM Movie 7. 

In the presence of renewal, topological strings and lattices thereof are stable for moderate activity, Fig.~\ref{fig:active_phase_diagram}b and SM Movies 8, 9.
In contrast to the case $\zeta=0$, however, active forces cause nematic point defects with charge $+1/2$ to align their axis with the topological string. Indeed, this is the only mechanically stable configuration for a defect pair subjected to a force couple.  
For a single string, extensile stress leads to outward-pointing forces stabilizing two nematic defects facing tail-to-tail, Fig.~\ref{fig:active_phase_diagram}b, top panel. The converse occurs for contractile stresses, which promote head-to-head configurations and thus asters. 
As a consequence, the far polar fields of $+1$ strings are vortices and asters, respectively. 
For large active stress, the self-propulsion of nematic point defects with charge $+1/2$ dominates  hydrostatic and Coulomb interactions leading to space-filling chaotic strings, SM Movie 10.

\textit{Discussion.---}In summary, we have shown that stable topological strings spontaneously emerge in nematopolar fluids with material renewal and identified the forces determining their length. Our analysis seems to be particularly relevant for the  cytoskeleton, where filamentous actin can organize into nematic or polar structures and material renewal results form continuous filament assembly and disassembly. We note that topological strings have the same orientational order as stress fibers in adherent animal cells. However, the actin density along stress fibers appears to be larger compared to the surrounding, which is in contrast to the strings studied in the present work. It thus remains to be seen whether the mechanisms studied here plays a role in generating stress fibers. 

In this study we used the simplest form of renewal. Cytoskeletal and other biological structures that fall into the class of nematopolar fluids might require more involved schemes~\cite{Hannezo:2015ba} or rely on biochemical regulation of renewal~\cite{Etienne-Manneville.2002}, which can lead to new patterns~\cite{barberi2023localized}. Further investigation of the impact of renewal regulation on topological strings and nematoplar structures in general is required to understand this mechanochemical link.

Finally, self-propelled particles (SPPs) provide another class of active system where mixed orientational order can emerge~\cite{baskaran2012self,bar2020self,Sousa.2025zt4y}. Also, coarse graining SPP dynamics automatically yields theories with mixed polarity~\cite{Peshkov.2012,bertin2015pre,bar2020self,spera2024nematic}. The consequences of mixed order in SPP remains largely unexplored, and it would  be particularly interesting to identify mechanisms that can stabilize topological defects.

\textit{Acknowledgments.---}We thank Birte Geerds, He Li, Yann Maggipinto, and Daniel Riveline for insightful discussion. This work was partially funded by Swiss National Science Foundation through grant number $200020\mathrm{E}\_219164$.
%\end{acknowledgments}

\newpage

\section{End matter}
\subsection{Non-dimensional units}
Following~\cite{dumoulin2025defect} we non-dimensionalize the dynamics we introduce the hydrodynamic lengthscale $\ell_{\rm h} \equiv \sqrt{\eta/\xi}$, where $\eta$ is the kinematic viscosity and $\xi$ the friction coefficient; an arbitrary time-scale $T$; a density scale is $\tilde \rho = T \eta \ell_{\rm h}^{-2}$. All non-dimensional quantities, which we here denote with the prime subscript, are obtained from the corresponding dimensional quantities as: $\tau' = \tau/T$, $\rho' = \rho \tilde\rho^{-1}$, $v_\alpha' = v_\alpha T \ell_h^{-1}$, $a' = a \tilde\rho^4 T \eta^{-1}$, $\alpha_X' = \alpha_X \ell_h^{-2} T \eta^{-1}$, $\beta_X' = \beta_X \ell_h^{-2} T \eta^{-1}$, $\chi' = \chi \ell_h^{-2} T \eta^{-1}$,  $\kappa_X' = \kappa_X \tilde\rho^2 T\eta^{-1} \ell_h^{-2}$, $\Gamma' = \Gamma \eta^{-1}$, $D' = D \ell_h^{-2} T$, $\zeta' = \zeta \tilde\rho T \eta^{-1}$.

\subsection{Numerical parameters for figures}
Throughout our work we fix the following non-dimensional parameters (we drop the prime index with respect to the previous section): $a = 1$, $\alpha_X = 0.1$, $\beta_X =0.1$, $\kappa_X = 10^{-4}$, $D = 10^{-4}$, $\Gamma = 1$. We vary $\rho_0$ between $0.4$---$1.2$, $\tau^{-1}$ between $0$---$10$, $\chi$ between $0.01$---$0.15$, $|\zeta|$ between $10^{-4}$---$3 \cdot 10^{-2}$. To ensure numerical stability, parameter exploration is limited by the constraint of positive compressibility $K(\rho_0) = \rho_0 \Pi'(\rho_0) > 0$ as detailed in the SM. 
In Figs.~\ref{fig:single-defect}a-c, \ref{fig:multiple-defect}-\ref{fig:active_phase_diagram} we use $L = 10$, $N=512$. In Figs.~\ref{fig:single-defect}d,~\ref{fig:pressure},~\ref{fig:charge_dependent} we use $L=20$, $N=2048$. Details on the initial conditions and  are given in the SM~\cite{supp}.

\subsection{Darcy-Brinkman flow}
\begin{figure}[htp!]
    \centering
    \includegraphics[width=1\linewidth]{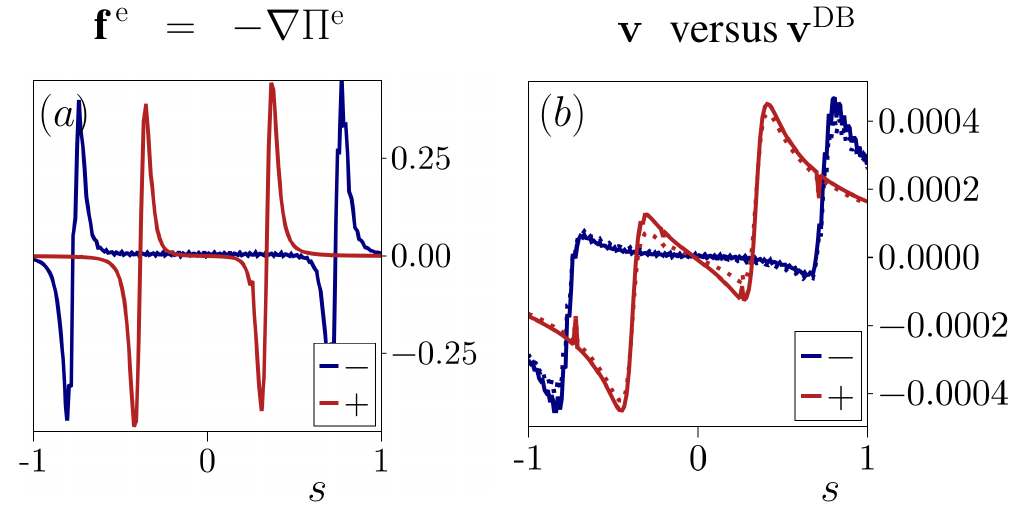}
    \caption{(a) Forces resulting from the effective pressure $\Pi^{\rm e}$ projected along the string axis, for positive and negative strings. (b) Comparison between local tangential velocity $v_\alpha$ (solid lines) and the Darcy-Brinkman flow $v^{\rm DB}_\alpha$ (dotted lines), for positive and negative strings. Same parameters as in Fig.~\ref{fig:pressure}a.}
    \label{fig:DarcyBrinkman}
\end{figure}

In Fig.~\ref{fig:DarcyBrinkman} we compare the measured flow $\bfv$ along a topological string with the Darcy-Brinkman flow $v^{\rm DB}_\alpha = - G_{\alpha\beta} \> \ast \partial_\beta \Pi^{\rm e}$ resulting from the effective pressure $\Pi^{\rm e}$. The Darcy-Brinkman propagator $G_{\alpha\beta}$ is the Green function of the operator $\cL_{\alpha\beta} v_\beta = \xi\, v_\alpha - \eta\,\partial_\gamma \left( \partial_\alpha v_\gamma + \partial_\gamma v_\alpha \right)$, computed in SM~\cite{supp}. In our units, $\xi = \eta = 1$. 
We observe that the flow field is captured very well by the Darcy-Brinkman flow generated by the effective pressure.

\bibliography{biblio.bib}

\end{document}